\documentclass[a4paper,11pt]{article}
\usepackage{graphicx,amssymb,bm,latexsym,color,epsf,cite,ulem}
\usepackage{hyperref}
\makeatletter
\@addtoreset{equation}{section}

\makeatother
\newcommand{\be}{\begin{equation}}
\newcommand{\ee}{\end{equation}}
\newcommand{\bea}{\begin{eqnarray}}
\newcommand{\eea}{\end{eqnarray}}
\newcommand{\vs}[1]{\vspace{#1 mm}}
\newcommand{\hs}[1]{\hspace{#1 mm}}
\renewcommand{\a}{\alpha}
\renewcommand{\b}{\beta}
\renewcommand{\c}{\gamma}

\renewcommand{\d}{\delta}

\newcommand{\s}{\sigma}
\renewcommand{\t}{\theta}

\newcommand{\la}{\lambda}
\newcommand{\pa}{\partial}

\newcommand{\nn}{\nonumber\\}
\newcommand{\p}[1]{(\ref{#1})}

\newcommand{\cO}{{\cal O}}
\newcommand{\cV}{{\cal V}}
\newcommand{\cL}{{\cal L}}

\newcommand{\bR}{\bar R}
\newcommand{\br}{\bar R}
\newcommand{\bg}{\bar g}

\newcommand{\bnabla}{\bar\nabla}

\begin{document}

\begin{flushright}
KU-TP 075 \\
\today
\end{flushright}

\begin{center}
{\Large\bf Towards the determination of the dimension\\
of the critical surface in asymptotically safe gravity}
\vs{10}

{\large
Kevin Falls,$^{a,b,}$\footnote{e-mail address: kfalls@sissa.it}
Nobuyoshi Ohta,$^{c,}$\footnote{e-mail address: ohtan@phys.kindai.ac.jp}
and
Roberto Percacci$^{a,b,}$\footnote{e-mail address: percacci@sissa.it}
} \\
\vs{10}
$^a${ International School for Advanced Studies, via Bonomea 265, 34136 Trieste, Italy}

$^b${ INFN, Sezione di Trieste, Italy}

$^c${Department of Physics, Kindai University,
Higashi-Osaka, Osaka 577-8502, Japan}

\vs{10}
%%%%%%%%%%%%%%%%%%%%%%%%%%%%%%%%
{\bf Abstract}
\end{center}

We compute the beta functions of Higher Derivative Gravity within the Functional Renormalization Group approach,
going beyond previously studied approximations.
We find that the presence of a nontrivial Newtonian coupling
induces, in addition to the free
fixed point of the one-loop approximation,
also two nontrivial fixed points,
of which one has the right signs to be free from tachyons.
Our results are consistent with earlier suggestions that the dimension of the critical surface for pure gravity is three.

%%%%%%%%%%%%%%%%%%%%%%
\section{Introduction}
%%%%%%%%%%%%%%%%%%%%%%

Higher Derivative Gravity (HDG) is the theory of gravity based on the metric as the carrier of degrees of freedom,
with an action containing terms of order zero, one and two in the curvature.
It contains both dimensionful couplings (the cosmological and Newton constant) and dimensionless ones
(the coefficients of the HD terms).
When treated perturbatively in the latter, it is renormalizable~\cite{stelle}, but not unitary.
Following some earlier attempts~\cite{julve,ft1},
its one-loop beta functions were correctly derived for the first time in~\cite{avrabar};
for more details and generalizations, see~\cite{deBerredoPeixoto:2003pj,deBerredoPeixoto:2004if}.
Depending on the signs of the couplings, the theory 
can be asymptotically free, but it has ghosts and/or tachyons.
There has been recently a revival of interest in this theory, 
and proposals to get around its problems in various
ways~\cite{Mannheim:2006rd,Salvio:2014soa,Salvio:2017qkx,Salvio:2018crh,Einhorn:2014gfa,AKKLR,smilga,holdom,donhdg,Anselmi:2018tmf,Donoghue:2019fcb}.

In the asymptotic safety approach to quantum gravity, one tries to construct a continuum limit around an interacting
fixed point (FP)~\cite{Weinberg}.
The main tool to investigate the gravitational renormalization group has been the Functional Renormalization Group Equation
(FRGE), as applied for the first time to gravity by Martin Reuter \cite{Reuter}.
It defines a flow on the theory space consisting of all diffeomorphism invariant functionals of the metric.
One expects that at an interacting gravitational FP, infinitely many gravitational couplings will be nonzero.
In spite of this complication, much evidence for the existence of such a FP has been collected so far \cite{perbook,rsbook}.

In the context of asymptotic safety, when one uses the FRGE,
there is never the need to postulate the form of the bare action
to be used in the path integral.
Instead, one directly calculates the flow of the effective action
as a function of an external ``coarse-graining'' scale,
or IR cutoff, $k$.
In this context, the action 
of HDG can be used as an ansatz for the running effective action.
We will call this the ``HDG truncation''.
It tracks the flow of the theory in a five-dimensional 
``theory space'' parametrized by the couplings:
$\cV$, $Z_N$, $\lambda$, $\xi$ and $\rho$, defined below.
The beta functions of HDG have been studied from this point of view in several papers.
They were obtained in a one-loop approximation to the FRGE in~\cite{Codello:2006in,niedermaier,OP2013,OPP2}.
In these calculations, the beta functions of the HD couplings 
are asymptotically free, 
in agreement with the old perturbative results,
but the flow of the dimensionful couplings looks very similar to the one of the Einstein-Hilbert truncation,
and exhibits a nontrivial FP for the cosmological and Newton constant.
To go beyond one loop, one has to keep terms involving
the beta functions in the r.h.s. of the flow equation,
and then solve these algebraic equations for the beta functions.
We highlight this process in Section 3.1.
This produces non-linearities that amount to resummations of
infinitely many loop diagrams.
This has been calculated in \cite{lauscher,bms} on a generic Einstein manifold, and a fully interacting FP was found,
but these calculations were limited to one or two, 
out of the three HD couplings.
This may seem to be sufficient, since one of the three couplings
is the coefficient of the Euler term, that does not contribute to the local dynamics.
Unfortunately, as we shall see in Sect.2.1,
on an Einstein manifold one computes the beta function of certain
linear combinations of the three couplings, and
it is actually impossible to identify the beta function 
of the two dynamically interesting ones: there is an unknown mixing with the beta function of the Euler term.
To compute the beta functions of all the independent couplings 
is the main task of this paper.

The main motivation for this is the determination of the dimension of the UV critical surface.
There is evidence from the $f(R)$ truncations that 
the scaling exponents at the nontrivial fixed point
are not too different from the classical ones,
so that couplings with positive mass dimension remain relevant
and couplings with negative mass dimension remain irrelevant
FP~\cite{CPR1,CPR2,FLNR2013,FKL2018,Falls:2018ylp}.
The marginal coupling of the $R^2$ term becomes relevant,
so altogether, in this truncation, 
the dimension of the critical surface seems to be three.
An attempt to include different tensor structures has been made in~\cite{FKL2018}, where actions of the form
$f_1(R_{\mu\nu}R^{\mu\nu})+R f_2(R_{\mu\nu}R^{\mu\nu})$
are studied, leading to the same conclusion.
A limitation of these calculations is that, 
on a spherical background, it is not possible to
properly disentangle independent couplings
with the same number of curvatures.
The case of Ricci tensor squared and scalar curvature squared actions
on an Einstein manifold, has already been cited above \cite{bms}.
While more general than spheres, Einstein manifolds are
still not general enough to distinguish all invariants.
With this limitation, it was found again that the dimension of the critical surface is three.
This suggests that some linear combination of the HD couplings
may be an irrelevant operator.
It seems possible, and even likely, that the dimension of the critical surface in pure gravity is determined entirely
by the fate of the HD couplings, since they are not expected to remain marginal at an interacting FP.\footnote{
So far the only indication that things could be more complicated comes from work in progress by Kluth and Litim
on actions of the form $f_1(R_{\mu\nu\rho\sigma}R^{\mu\nu\rho\sigma})+R f_2(R_{\mu\nu\rho\sigma}R^{\mu\nu\rho\sigma})$,
where a term cubic in curvature seems to become relevant~\cite{kluth}.}
We find that of the three dimensionless couplings,
one becomes relevant, one irrelevant and one --
the coefficient of the Euler term -- remains marginal.
The beta function of the Euler term is related to the $a$-function.
The $a$-theorem states that when two fixed points are joined
by an RG trajectory, the value of $a$ at the IR fixed point is lower
than the one at the UV fixed point.
We find some evidence that this may hold also in gravity.

In the present paper we try to shed some light on these issues by computing the beta functions of all the HD couplings
beyond the one-loop approximation, taking the anomalous 
dimensions into account.
We shall do this by using the ``Universal RG Machine''~
to compute the r.h.s. of the FRGE on an arbitrary background.
This is a technique based on non-diagonal heat kernel coefficients that can be used to evaluate functional traces
involving covariant derivatives acting on a function of a Laplacian.
The Universal RG Machine has been introduced, and applied to the Einstein-Hilbert truncation, in \cite{BGMS2010}.
Later it was used to calculate the one-loop beta functions
in HDG \cite{SGRZ}.
Technical details are given in \cite{GSZ2011}.
Here we bring that program one step forward by
evaluating the full beta functions of HDG,
including the anomalous dimensions.
The main steps of the calculation are outlined in Sect.2,
and in Sect.3 we describe the results.
We find three fixed points, of which one has vanishing higher derivative couplings, while the others are fully interacting.
In principle, any of these could be a viable UV fixed point.
To have a viable theory, one would also have to prove unitarity.
For the first of these fixed points, one could apply
the arguments developed in perturbation theory
\cite{Mannheim:2006rd,Salvio:2014soa,Salvio:2017qkx,Salvio:2018crh,Einhorn:2014gfa,AKKLR,smilga,holdom,donhdg,Anselmi:2018tmf,Donoghue:2019fcb}.
For the remaining ones, the issue is more involved
and will require a detailed study of the two point function.

%%%%%%%%%%%%%%%%%%%%%%%%%%%%%%%%%%%%%%%%%%%%%%%%%%%
\section{Beta functions}
%%%%%%%%%%%%%%%%%%%%%%%%%%%%%%%%%%%%%%%%%%%%%%%%%%%

\subsection{Why Einstein backgrounds are not enough}

Let us momentarily concentrate on the HD terms, that we can write
as $\cL_{HD}=\a R^2+\b R_{\mu\nu}^2 + \c R_{\mu\nu\rho\la}^2$.
Due to the fact that the Gauss--Bonnet combination
$E = R_{\mu\nu\a\b}^2- 4 R_{\mu\nu}^2+ R^2$
is topological, one of these couplings is uninteresting
as far as local dynamics is concerned.
It is therefore more meaningful to write the Lagrangian as
\be
\cL_{HD}=\frac{1}{2\la} C^2 + \frac{1}{\xi} R^2 - \frac{1}{\rho} E
\ee
where
\be
\frac{1}{\xi}= \frac{3\alpha+\beta+\gamma}{3}\ ,\quad
\frac{1}{2\lambda} = \frac{\beta+4\gamma}{2}\ ,\quad
-\frac{1}{\rho} = -\frac{\beta+2\gamma}{2}\ .
\ee
and
$C^2 = R_{\mu\nu\a\b}^2-2R_{\mu\nu}^2+ \frac13 R^2$
is the square of the Weyl tensor.
We are mainly interested in the beta functions of $\lambda$ and $\xi$.
Calculations are simpler on an Einstein background.
In this case $E=R_{\mu\nu\rho\sigma}R^{\mu\nu\rho\sigma}$
and $C^2=R_{\mu\nu\rho\sigma}R^{\mu\nu\rho\sigma}-R^2/6$, so
\be
\cL_{HD}=\left(\frac{1}{\xi}-\frac{1}{12\lambda}\right)R^2
+\left(\frac{1}{2\lambda}-\frac{1}{\rho}\right) E 
%R_{\mu\nu\rho\sigma}R^{\mu\nu\rho\sigma}
\ .
\ee
This implies that if we expand the r.h.s. of the functional RG equation
on an Einstein background, and we interpret the coefficients
of $R^2$ and $E=R_{\mu\nu\rho\sigma}R^{\mu\nu\rho\sigma}$
as beta functions, we can read off
the beta functions of two combinations of $\lambda$, $\xi$, $\rho$
but we are unable to unambiguously identify $\beta_\lambda$ and $\beta_\xi$.
To do this, we need an additional independent equation,
that in turn requires a more general background.
This is what we do in this paper.

All calculations will be based on the Euclidean action
\bea
S&=&\int d^4 x \sqrt{g} \Big[\cV-Z_N R+\cL_{HD}
\Big] ,
\label{action}
\eea
where $Z_N=\frac{1}{16\pi G}$, $G$ being Newton's constant,
$\cV=2\Lambda Z_N$ and $\Lambda$ is the cosmological constant.
Sometimes we shall use the combinations
\bea
\omega \equiv -\frac{3\la}{\xi},~~~
\theta \equiv \frac{\la}{\rho}\ .
\eea

\subsection{Remark on the topological term}

Before embarking in calculations, we can make a general remark
on the Gauss-Bonnet term, that actually holds
{\it independently of the truncation}.
Due to the topological character of the term $E$,
its coefficient $1/\rho$ does not appear in the Hessian
and therefore does not appear in the r.h.s. of the flow equation.
Thus the beta function of $\rho$ must have the form
\be
\beta_\rho=-\frac{1}{16\pi^2}a\rho^2\ .
\label{betarho}
\ee
where $a$ is a function of all the other couplings,
but not of $\rho$ itself.
In the search of a fixed point one can solve first the equations
of all the other couplings, which are also independent of $\rho$.
When these fixed point values are inserted in (\ref{betarho}),
$a$ becomes just a number.
The UV behavior of $\rho$ is determined by the value of this number.
If $a=0$, $\rho$ could reach any value in the UV.
If $a>0$ ($a<0$), when all other couplings are very close to
a fixed point, it will run logarithmically to zero
from above (below).

\subsection{Expansion and gauge fixing}

We split the metric $g_{\mu\nu} = \bg_{\mu\nu} + h_{\mu\nu}$, 
where $\bg_{\mu\nu}$ is an arbitrary background.
For details of the expansion of the action, we refer to \cite{OP2013}.
The gauge-fixing and ghost action can be written
\bea
{\cal L}_{GF+FP}/\sqrt{\bg}
= -\frac{1}{2a} \chi_\mu Y^{\mu\nu} \chi_\nu + i Z_{gh} \bar c^\mu \Delta^{(gh)}_{\mu\nu} c^\nu
 + \frac{1}{2} Z_Y b_\mu Y^{\mu\nu} b_\nu  +  Z_Y \bar{\zeta}_\mu  Y^{\mu\nu} \zeta_\nu
\label{gfgh1}
\eea
where $\bar c_\mu$, $c_\mu$ are complex ghosts, $b_\mu$ is a real commuting field, $\bar{\zeta}_\mu$, $\zeta_\mu$ are complex anti-commuting fields and
\bea
\chi_\mu &\equiv& \bnabla^\la h_{\la\mu} + b \bnabla_\mu h\ , \nn
\Delta^{(gh)}_{\mu\nu} &\equiv & g_{\mu\nu} \bnabla^2 +(2b+1)\bnabla_\mu \bnabla_\nu +\br_{\mu\nu}\ , \nn
Y_{\mu\nu} &\equiv& \bg_{\mu\nu} \bnabla^2+ c \bnabla_\mu \bnabla_\nu - f \bnabla_\nu \bnabla_\mu\ .
\eea
where $a$, $b$, $c$ and $f$ are gauge parameters.
There is some freedom in how we choose the wave function renormalisations $Z_{gh}$ and $Z_{Y}$
since they can be rescaled while keeping $Z_{gh}^2 Z_Y=1/a$ 
fixed without  affecting the path integral.
In our calculations we fix 
\be
Z_{gh}=1\ ,\quad Z_Y=1/a
\label{zgh}
\ee
We make the usual gauge choice
\bea
a= %\frac{1}{\b+4\c}=
\lambda\ ,~~~
b= %\frac{4\a+\b}{4(\c-\a)}, ~~~
-\frac{1+4\omega}{4+4\omega}\ ,~~~
c = \frac23(1+\omega)\ ,~~~
%\frac{2(\c-\a)}{\b+4\c}-1
f=1\ ,
\label{gpar}
\eea
leading to a minimal fourth order operator for the fluctuations.
The operators in \p{gfgh1} are then
\bea
\Delta^{(gh)}_{\mu\nu} &\equiv & 
g_{\mu\nu} \bnabla^2 -\sigma_{gh}\bnabla_\mu \bnabla_\nu +\br_{\mu\nu}, \nn
Y_{\mu\nu} &\equiv& \bg_{\mu\nu} \bnabla^2 -\sigma_Y\bnabla_\mu \bnabla_\nu - R_{\mu\nu},
\eea
with
\bea
\sigma_{gh}=-1-2b
=-\frac{1-2\omega}{2(1+\omega)}
\ ;\qquad
\sigma_Y=1-2\frac{\gamma-\alpha}{\beta+4\gamma}=\frac{1-2\omega}{3}\ .
\label{sigmas}
\eea
We note that the cancellation between unphysical degrees of freedom
becomes exact in the ``Landau gauge'' limit $a\to 0$,
which happens to be satisfied in the
asymptotically free regime.

Then, the quadratic terms in the action can be written in the form~\cite{OP2013}
\bea
{\cal L}^{(2)}= h_{\mu\nu}K^{\mu\nu\rho\sigma}
\cO_{\rho\sigma}{}^{\alpha\beta} h_{\a\b},
\eea
where the operator $\cO$ is
\bea
{\cal O}=\Delta^2 + V_{\rho\la} \bnabla^\rho \bnabla^\la +U\ .
\label{hami}
\eea
with $\Delta=-\bnabla^2$,
$U=K^{-1}W$ and we write
\bea
K = \frac{\b+4\c}{4} \Big(\mathbb{I}
+\frac{4\a+\b}{\c-\a}\mathbb{P} \Big)\ ,
\qquad
K^{-1}= \frac{4}{\b+4\c}
\Big(\mathbb{I}
-\frac{4\a+\b}{3\a+\b+\c}\mathbb{P}  \Big),
\label{kkinv}
\eea
where
$\mathbb{I}$ is the identity in the space of symmetric tensor
and $\mathbb{P}$ is a projector
\bea
\mathbb{I}_{\mu\nu,\a\b}\equiv \d_{\mu\nu,\a\b}=\frac{1}{2} (\bg_{\mu\a} \bg_{\nu\b}+\bg_{\mu\b} \bg_{\nu\a})\ ,\quad
\mathbb{P}^{\mu\nu}{}_{\rho\sigma} \equiv P^{\mu\nu}{}_{\rho\sigma}=\frac{1}{4}\bg^{\mu\nu}
\bg_{\rho\sigma}\ .
\eea
The coefficients $V_{\rho\lambda}$ and $U$ are functions
of the curvatures, $\cV$ and $Z_N$, for whose form
we refer again to \cite{OP2013}.

The ``beta functional'' of the theory is the sum of three
contributions coming from gravitons, ghosts and the new ghost $b_\mu$:
\bea
\dot\Gamma_k &=&T_g+T_{gh}+T_Y\ .
\eea
In order to write these terms more explicitly, we have to choose
a cutoff for each of them.
For a one-loop calculation, where the couplings in the r.h.s. of the equation are treated as fixed,
it was most convenient to think of the cutoff as a function of the whole
operator ${\cal O}$, $\Delta_{gh}$ or $Y$ respectively
(so-called type III cutoff).
In this paper we will not ignore the running of the couplings that may be present in the cutoff,
so it is best to minimize their presence. 
This is achieved by choosing the cutoff to be a function of $\Delta$ only (so-called type I cutoff).
The one-loop calculation with this cutoff has been
done before in \cite{SGRZ}.

%%%%%%%%%%%%%%%%%%%%%%%%%%%%%%%%%%%%%%%
\subsection{Graviton contribution}
%%%%%%%%%%%%%%%%%%%%%%%%%%%%%%%%%%%%%%%

We choose the graviton cutoff to have the form
${\cal R}=K R_k(\Delta^2)$,
where $R_k(\Delta^2)=(k^4-\Delta^2)\theta(k^4-\Delta^2)$
and we define as usual
$P_k(\Delta^2)=\Delta^2+R_k(\Delta^2)=k^4\t(k^4-\Delta^2)$.
Note that it is convenient to view $R_k$ as a function of $\Delta^2$,
although of course one could also view it as a function of $\Delta$.
Then, writing the kinetic operator as $\Delta^2+V+U$,
the graviton contribution to the FRGE is
\be
T_g=\frac{1}{2}\mbox{Tr}\frac{\pa_t[K R_k(\Delta^2)]}{K [{\cal O}+R_k(\Delta^2)]}
=\frac{1}{2}\mbox{Tr}\frac{\pa_t R_k(\Delta^2)+\eta_K R_k(\Delta^2)}{P_k(\Delta^2)+V+U}\ ,
\label{frgeg}
\ee
where we defined 
\be
\eta_K = K^{-1}\frac{dK}{dt} .
\ee
Note that $\eta_K$ is a tensor. 
From (\ref{kkinv}) we find
\bea
\eta_K=\eta_1\mathbb{I}+\eta_P\mathbb{P},
\label{etaK}
\eea
where
\bea
\eta_1=-\frac{\dot\lambda}{\lambda}\ ,\qquad
\eta_P=
-\frac{\xi\dot\lambda-\lambda\dot\xi}{\lambda(3\lambda-\xi)}\ ,
\label{etap}
\eea
We divide $V$ and $U$ into various terms:
$V=V_0+V_1$ and $U=U_0+U_1+U_2$,
where the subscript counts the power of curvature,
and the remaining dimension is carried either by $\cV$ or $Z_N$:
$$
V_0\sim Z_N\nabla\nabla\ ;\qquad
V_1\sim R\nabla\nabla\ ;\qquad
U_0\sim \cV\ ;\qquad
U_1\sim Z_N R\ ;\qquad
U_2\sim R^2\ .
$$
We now have to decide how to expand the fraction in (\ref{frgeg}).
Since we want to compute the beta functions of all the couplings in
(\ref{action}), we need to expand to second order in curvatures.
It would be natural to assume that $\sqrt{\cV}\sim Z_N\sim R$
(which implies also $\Lambda\sim R$),
but such an expansion would miss important features,
as we shall discuss below.
It is possible without too much effort to keep the {\it full} dependence on $\cV$, and we shall do so. 
We will therefore not expand in $U_0$.
It is much harder to keep all dependence on $Z_N$,
therefore we will expand in $V_0$, $V_1$, $U_1$ and $U_2$,
to first order in $Z_N/k^2$, independently of curvatures.\footnote{Note that we wrote $\cV=2Z_N\Lambda$ and treated
$\Lambda$ as an independent coupling, the expansion in $Z_N$
would also entail and expansion in $\Lambda$. 
This is not what we do here.}
This corresponds to considering a trans-Planckian regime.
If one considers the Einstein-Hilbert part of the action,
it correspond to a {\it strong gravity} expansion.
See \cite{Niedermaier:2019iqk} for a recent discussion.
Keeping only terms up to linear order in $Z_N$ we thus have
to evaluate:
\bea
T^{\rm grav}&=&\frac12 \mbox{Tr}\left[ 
\frac{\pa_t R_k(\Delta)+\eta_K R_k(\Delta)}{P_k(\Delta)+U_0}
\left(1-\frac{1}{P_k(\Delta)+U_0}(V_0+V_1+U_1+U_2)
\right.\right. \nn
&& \hs{-10} \left.\left.
+\frac{1}{P_k(\Delta)+U_0}V_0 \,
\frac{1}{P_k(\Delta)+U_0}V_1
+\frac{1}{P_k(\Delta)+U_0}V_1 \,
\frac{1}{P_k(\Delta)+U_0}V_0  
\right.\right. \nn
&& \hs{-10} \left.\left.
+\frac{2V_0 U_2}{(P_k(\Delta)+U_0)^2}
+\frac{V_1^2}{(P_k(\Delta)+U_0)^2} 
+\frac{2V_1 U_1}{(P_k(\Delta)+U_0)^2} 
+\frac{3V_0 V_1^2}{(P_k(\Delta)+U_0)^3} 
\right)\right]
\ .
\label{tgrav}
\eea
In the last line we have written the terms only in a schematic way,
without paying attention to their order:
to be precise one has to write out several terms where the
projectors $\mathbb{P}$ appear in different positions.
(For details, we refer the reader to the ancillary file on the arXiv page.)

%%%%%%%%%%%%%%%%%%%
\subsection{Ghost contribution}
%%%%%%%%%%%%%%%%%%%

To some extent, it is possible to treat $\Delta_{gh}$ and $Y$
together.
Both operators are non-minimal, and of the form 
$\Delta \delta_\mu^\nu+\sigma \bnabla_\mu\bnabla^\nu+B_\mu^\nu$
(note the overall sign is reversed),
where $\sigma$ is a constant defined in \p{sigmas} and $B_\mu^\nu=s \bR_\mu^\nu$,
where $s=-1$ for $\Delta_{gh}$ and $s=1$ for $Y$.
In the standard one-loop calculations, one can use the known
heat kernel coefficients for this type of operators.
In contrast to \cite{Codello:2006in,niedermaier,OP2013} and
coherently with the treatment of gravitons,
we use a type I cutoff also for the ghosts.
This type of cutoff for ghosts had been used before in \cite{SGRZ}.
The novelty of our calculation is that we also
take into account the contributions due to the anomalous dimensions
\be
\eta_{gh}=0\ ,\qquad
\eta_Y=-\beta_\lambda/\lambda\ .
\label{etaY}
\ee
The type I cutoff has the form\footnote{We observe that
the calculation of the ghost contributions is considerably simpler
with a so-called type-II cutoff
${\cal R}^\mu_{k\,\nu}=Z\delta^\mu_\nu R_k(\Delta+B)$.
The use of this alternative scheme for the ghosts
would lead to only small quantitative differences
in the final results for the fixed points and
we shall not discuss this in detail.}
\be
{\cal R}^\mu_{k\,\nu}=Z\delta_\mu^\nu R_k(\Delta),
\ee
where $Z$ is given by (\ref{zgh},\ref{gpar}).
Adding the cutoff, the kinetic operator (aside from the factor $Z$)
becomes
$P_k(\Delta) \delta_\mu^\nu+\sigma \bnabla_\mu\bnabla^\nu+B_\mu^\nu$.
In the flow equation one needs the inverse of this operator.
We refer to \cite{SGRZ} for some technical details.
The evaluation of the traces to second order in curvatures
is rather laborious.
In the end we arrive at the following
\bea
&& \hs{-10}T_{gh}= -\frac{1}{(4\pi)^2} \int d^4 x \sqrt{\bg}
\Bigg\{ 
\left[ 3-\frac{2}{\s_{gh}}-\frac{2}{\s_{gh}^2}\log(1-\s_{gh})
\right]k^4
\nn
&& -\frac{1}{12\s_{gh}^2}\left[\frac{3\s_{gh}(2+\s_{gh}(7-5\s_{gh}))}{\s_{gh}-1}
-2(3-2\s_{gh}) \log(1-\s_{gh})\right]k^2 \br
\nn
&& -\frac{11}{90}\br_{\mu\nu\rho\la}^2 
+\frac{43-2\s_{gh}(13+\s_{gh})}{45(1-\s_{gh})^2} \br_{\mu\nu}^2
+ \left[\frac{5}{18}+\frac{1}{6(1-\s_{gh})^2}\right] \br^2
 \Bigg\} .
 \label{tgh}
\eea
Note the appearance of $\log(1-\sigma_{gh})=-\log(2(1+\omega)/3)$,
which forces us to consider only the domain $\omega>-1$.
For $Y$:
\bea
&& \hs{-10}T_Y=-\frac{1}{2} \frac{1}{(4\pi)^2} \int d^4 x \sqrt{\bg}
\Bigg\{ 
\Bigg[ 3-\frac{2}{\s_Y}-\frac{2}{\s_Y^2}\log(1-\s_Y)
\nn
&& 
\qquad\qquad\qquad\qquad\qquad\qquad
+\eta_Y\left( \frac{2-\s_Y+\s_Y^2}{2\s_Y^2}
+\frac{(1-\s_Y)}{\s_Y^3}\log(1-\s_Y)\right)\Bigg]k^4
\nn
&&\!\!\!\!\!\!\!
+\left[-\frac{2+\s_Y}{4\s_Y}
-\frac{3+2\s_Y}{6\s_Y^2}\log(1-\s_Y)
+\eta_Y\Big(\frac{6-\s_Y}{12\s_Y^2}
+\frac{3-2\s_Y-\s_Y^2}{6\s_Y^3} \log(1-\s_Y) \Big)\right]k^2 \br
\nn
&& 
\!\!\!\!\!\!\!
-\frac{11}{90}\Big(1+\frac{\eta_Y}{2} \Big) \br_{\mu\nu\rho\la}^2   
%\nn
%&& 
\!+\!\left[\frac{43}{45}
+\eta_Y \Big(\frac{20\!-\!20\s_Y\!-\!39\s_Y^2\!+\!29\s_Y^3}{120\s_Y^2(\s_Y-1)}
-\frac{1\!-\!\s_Y\!-\!2\s_Y^2}{12\s_Y^3}\log(1\!-\!\s_Y)
\Big)\right] \br_{\mu\nu}^2
\nn
&& \!\!\!\!\!\!\!
- \left[ \frac{2}{9} 
+\eta_Y\Big(\frac{4+\s_Y^2+\s_Y^3-3\s_Y^4}{48(-1+\s_Y)\s_Y^2}
-\frac{2-\s_Y-2\s_Y^2}{24\s_Y^3}\log(1-\s)\Big)
\right] \br^2
\Bigg\} ,
\label{tY}
\eea
Both agree with \cite{SGRZ} if we put $\eta_Y=0$.

%%%%%%%%%%%%%%%%%%%%%%%%%%%%%%%%%%%%%%%%%%%
%%%%%%%%%%%%%%%%%%%%%%%%%%%%%%%%%%%%%%%%%%%
\section{Results}
%%%%%%%%%%%%%%%%%%%%%%%%%%%%%%%%%%%%%%%%%%%
%%%%%%%%%%%%%%%%%%%%%%%%%%%%%%%%%%%%%%%%%%%

\subsection{Beta functions}

For the study of the flow, the dimensionful couplings
$\cV$ and $Z_N$ have to be replaced by their
dimensionless counterparts $\tilde\cV=\cV/k^4$ and 
$\tilde Z_N=Z_N/k^2$, or the related quantities
$\tilde G=Gk^2$, $\tilde\Lambda=\Lambda/k^2$.
The beta functions are too complicated to be written here
(they are given in a Mathematica notebook \cite{kevin_falls_2020_4017671}),
but they simplify in two cases.
Expanding for small $\lambda$ we obtain the universal one-loop beta functions 
\bea
\beta_\lambda & =& -\frac{133 \lambda ^2}{160 \pi ^2}+O\left(\lambda ^3\right) 
\\
\beta_\omega &=& - \frac{\lambda  \left(200 \omega ^2+1098 \omega +25\right)}{960 \pi ^2}   +O\left(\lambda^2\right) 
\\
\beta_\theta &=& \frac{7(56-171\theta)}{1440\pi ^2}\lambda+O\left(\lambda^2\right) 
\eea
while the non-universal beta functions for $\tilde G$ and $\tilde\Lambda$ agree with those found in the one-loop calculation \cite{SGRZ} at $\lambda =0$.
Explicitly they are given by
\bea
\beta_{\tilde G} &=&   2 \tilde G+\tilde G^2 
\left[-\frac{c_1}{72 \pi  (1-2 \omega )}+\frac{c_2
   \log \left(\frac{2 (1+\omega )}{3}\right)}{12 \pi  (1-2 \omega )^2}\right] +  O\left(\lambda \right) 
   \label{oneloopg}
\\
\beta_{\tilde \Lambda} &=& -2 \tilde\Lambda 
+\frac{\tilde G}{72 \pi} \left[\frac{c_3+\tilde\Lambda  c_4 }{1-2 \omega
   }+\frac{6 \left(c_5+\tilde\Lambda  c_6 \right) \log
   \left(\frac{2 (1+\omega )}{3}\right)}{(1-2 \omega )^2}\right] +  O\left(\lambda \right) 
\eea
with the coefficients
$c_1= 35-2 \omega  (109+176 \omega )$, 
$c_2 = 65+4 \omega  (7+2 \omega )$, 
$c_3 = 162-540\omega$,
$c_4= -35+218\omega+352\omega^2$,
$c_5 = 6-96\omega-48\omega^2$,
$c_6= 65+28\omega+8\omega^2$.

Our calculation differs from one-loop calculations
in that we take into account the anomalous dimensions.
For example, we see $\eta_Y$ appearing explicitly in (\ref{tY}),
which gives contributions to the beta functions of all the couplings.
Equation (\ref{etaY}) tells us that $\eta_Y$ is proportional to
$\beta_\lambda$.
Thus, comparing the terms proportional to $C^2$ on both sides
of the FRGE, we obtain a relation of the form
$\beta_i=B_i+C_{ij}\beta_j$.
At one loop one just keeps $\beta_i=B_i$.
Solving  the algebraic equations gives beta functions
that contain contributions with arbitrarily high loop order.

However, from the definitions,
the anomalous dimensions
at a fixed point are known a priori to be
\be
\eta_1=0\ ;\qquad
\eta_P=0\ ;\qquad
\eta_Y=0\ .
\label{andimfp}
\ee
So, in the search of fixed points,
one can use simplified beta functions where
these values are used:
the full expressions for the anomalous dimensions 
are only needed when one calculates the scaling exponents.
It is easy to see that if we had assumed that
all terms in $V$ and $U$ are of the same order,
namely $\sqrt\cV\sim Z_N\sim R$,
then all the terms containing $V_0$ and $U_1$
would not contribute to the beta functions of
$\lambda$, $\xi$ and $\rho$.
Therefore, these beta functions would not contain $Z_N$
and would be exactly the same
as in 
\cite{SGRZ}.
This is why it is important to keep the expansion in $Z_N$
separate from the expansion in $R$.\footnote{It would obviously be even better not to expand
in $Z_N$ at all, but this would be technically much more challenging.}

Even the simplified beta functions with (\ref{andimfp})
are too complicated to be reported in detail.
However, we shall see a posteriori that $\tilde\cV$
is very small at fixed points.
If we put $\tilde\cV=0$, the equations for the remaining variables
become simple enough:
\bea
\beta_\lambda&=&-\frac{133}{160\pi^2}\lambda^2+\tilde Z_N\lambda^3\frac{251\xi-58\lambda}{120\pi^2\xi}
\label{blam}
\\
\beta_\xi&=&-\frac{5(72\lambda^2-36\lambda\xi+\xi^2)}{576\pi^2}+\tilde Z_N\frac{9720\lambda^3-1980\lambda^2\xi
+489\lambda\xi^2-14\xi^3}{6480\pi^2}
\\
\beta_\rho&=&-\frac{49}{180\pi^2}\rho^2+\tilde Z_N\lambda\rho^2\frac{233\xi-58\lambda}{240\pi^2\xi}
\label{brho}
\\
\beta_{\tilde Z_N}&=&\left(-2+\frac{(30\lambda-\xi)(4\lambda+\xi)}{192\pi^2\xi}\right)\tilde Z_N
+\frac{-3168\lambda^2+654\lambda\xi+35\xi^2}{1152\pi^2\xi(6\lambda+\xi)}\nonumber\\
&&\qquad\qquad\qquad-\frac{72\lambda^2-84\lambda\xi+65\xi^2}{192\pi^2(6\lambda+\xi)^2}
\log\left(\frac23-\frac{2\lambda}{\xi}\right)\ .
\label{bZ}
\eea

\subsection{Fixed points}

Now we recall that already in the one-loop calculation,
the beta functions of $\tilde Z_N$ (and also $\tilde\cV$) have a nontrivial fixed point.
This nonzero value of $Z_N$ enters in the beta functions
of (\ref{blam}-\ref{brho}) in such a way that besides
the asymptotically free fixed point, there are now two 
(and only two) new ones. Their coordinates are
given in Table 1.
\begin{table}[h]
\begin{center}
\begin{tabular}{|c|r|r|r|r|r|r|}
\hline
 & $\la_*$ & $\xi_*$ & $\rho_*$ & $\omega_*$ & $\tilde Z_{N*}$  & $\tilde G_*$  \\
\hline
FP$_1$ & 0 & 0 & 0 & $-0.02286$ & 0.00833 &   2.388   \\
\hline
FP$_2$ & 29.26 & $-220.2$ & 0 & 0.4040 & 0.01318 &   1.509   \\
\hline
FP$_3$ & 52.61 & 1672 & 0 & $-0.0944$ & 0.00761 &   2.614 \\
\hline
\end{tabular}
\end{center}
\caption{Fixed points in the approximation $\tilde\cV=0$.}
\label{t2}
\end{table}

The first fixed point is found also in the one-loop approximation,
and it is a non-trivial fact that it persists also when $\tilde Z_N$
is present in the beta functions of $\lambda$ and $\xi$.\footnote{Actually, this fixed point is best studied
using the variable $\omega$ instead of $\xi$. It corresponds to letting $\lambda$
and $\xi$ go to zero with a particular ratio,
and is different from setting e.g. first $\lambda=0$
and then $\xi=0$.}
Note that in the one-loop approximation there is also
another fixed point with $\lambda=\xi=0$, $\omega=-5.467$,
which however is excluded by our condition $\omega>-1$ 
(otherwise it gives a complex $\tilde Z_N$).
The remaining two fixed points are ``fully interacting''.
It is worth noting that if we treat $\tilde Z_{N}$ as an external parameter in the beta functions
of $\lambda$ and $\xi$, we find that $\lambda_*$ and $\xi_*$
go to infinity for $\tilde Z_{N}\to 0$.
\footnote{and to zero for $\tilde Z_{N}\to\infty$, but this
is outside the domain of our approximation.}

We then come to the solution of the full flow equations,
where we take into account also the running of $\tilde\cV$.
There are now more fixed points, and we report in Table 2 the properties
of the most interesting ones.

\begin{table}[h]
\begin{center}
\begin{tabular}{|c|r|r|r|r|r|r|r|r|r|}
\hline
 & $\la_*$ & $\xi_*$ & $\rho_*$ &  $\omega_*$ & $\tilde Z_{N*}$ & $\tilde\cV_*$ & $\tilde G_*$ & $\tilde\Lambda_*$ & $a$ \\
\hline
FP$_1$ & 0 & 0 & 0 & $-0.02286$ & 0.00833 & 0.006487 & 2.388 &  0.3894 & 4.356 \\
\hline
FP$_2$ & 24.91 & $-287.1$ & 0 &0.2603 & 0.01635 & 0.004575 & 1.217 & 0.1399 & $-2.741$ \\
\hline
FP$_3$ & 28.24 & 175.6 & 0 & $-0.4825$ & 0.01499 & 0.006928 & 1.327 & 0.2310 & $-3.566$ \\
\hline
FP$_4$ & 0 & $-312.2$ & 0 & 0 & 0.009222 & 0.006092 & 2.157 & 0.3303 & $4.357$ \\
\hline
\end{tabular}
\end{center}
\caption{Selected fixed points including $\tilde\cV$.}
\label{t2}
\end{table}

We see that in all cases the fixed point value of $\tilde\cV$ is very small,
justifying the earlier approximation $\cV=0$.
In fact, by considering only the beta functions of $\lambda$, $\xi$
and $\tilde Z_N$, and treating $\tilde\cV$ as a parameter,
and letting this parameter vary between zero and $0.004575$,
or $0.006928$, we can see that FP$_2$ and FP$_3$
change continuously from the values of Table 1 to those of Table 2.
We may thus identify the first three fixed points of Table 2 
with those of Table 1.

There are several other fixed points with $\lambda=0$,
of which FP$_4$ is a representative example.
We list it here for reasons that will become clear later.
There may also exist other non-trivial fixed points with $\lambda\not=0$,
but this would require a more extensive numerical search that
we have not undertaken.
Besides, these fixed points are probably artifacts of the 
truncation, as are known to occur in other similar cases.

We note that also $\tilde Z_{N*}$ is small,
and this justifies {\it a posteriori} the expansion in $\tilde Z_N$
that we used throughout our calculations.
If we change variable from $\tilde Z_N$ to $\tilde G_N$
and set $\lambda=0$, then as seen from (\ref{oneloopg})
there is a fixed point at $\tilde G=0$.
On the other hand, if we first set $\tilde G=0$,
there is no acceptable fixed point for the dimensionless couplings.
In any case, since we have expanded in $\tilde Z_N$,
any result near $\tilde G=0$ is unreliable.
This is unfortunate, because it means that we cannot check whether there exist a RG trajectory joining one of the fixed points listed above 
to the standard weak gravity regime in the IR.

\subsection{Scaling exponents}

If we rescale the fluctution field $h_{\mu\nu}$ by a factor
$\sqrt\lambda$, so that the prefactor of its kinetic term is canonical,
the fixed point FP$_1$ is seen to be a Gaussian fixed point,
and indeed we find that the scaling exponents are given by
the canonical dimensions: $4$, $2$, $0$, $0$, $0$.
The scaling exponents of FP$_2$, listed from more to less relevant, are
$$
\theta_{1,2}=2.35191 \pm 1.67715 i\ ,\quad 
\theta_3=1.76672\ ,\quad
\theta_4=0\ ,\quad
\theta_5=-3.20030\,,
$$
while those of FP$_3$ are
$$
\theta_{1,2}=2.03270 \pm 1.52155 i\ ,\quad 
\theta_3=1.23742\ ,\quad
\theta_4=0\ ,\quad
\theta_5=-5.27685\,.
$$
The marginal coupling is $\rho$, the (inverse of the) coefficient
of the topological term.
At the non-Gaussian fixed points, we find
$\beta_\rho=A\rho^2$ with $A=0.01736$ at FP$_2$
and $A=0.02258$ at FP$_3$.
Thus, at both fixed points, $\rho$ is marginally relevant
when it is negative and marginally irrelevant
when it is positive.
We thus arrive at the conclusion that also in the present 
approximation, the dimension of the critical surface
of pure gravity is three, up to the marginal topological term.

\subsection{The $a$-function}

The beta function of $\rho$ is given by (\ref{betarho}).
In an ordinary CFT, the coefficient $a$ 
appears in the trace anomaly as
\be
\langle T^\mu{}_\mu\rangle=
\frac{1}{16\pi^2}(c C_{\mu\nu\rho\sigma}C^{\mu\nu\rho\sigma}-a E)\ .
\ee
For example, for a free theory with $N_S$ scalars,
$N_f$ Dirac fields and $N_V$ gauge fields,
\be
a=\frac{1}{360}(N_S+11 N_f+62 N_V)\ ,\quad
c=\frac{1}{120}(N_S+6 N_f+12 N_V)\ .
\label{matter}
\ee
According to the $a$-theorem, 
if there is a RG trajectory joining two fixed points,
$a$ is higher at the UV fixed point \cite{Cardy:1988cwa,Komargodski:2011vj,Shore:2016xor}.
This accords to the intuition that $a$ is a measure of the number
of degrees of freedom of the theory.
There is no known $a$-theorem for gravity.
However, we can view our calculation as a quantum field theory
in a curved background, and from this point of view the theorem should
be applicable.\footnote{Similar calculations involving gravity have been
reported in \cite{Antoniadis:1992xu,Antoniadis:1996pb}.}
At FP$_1$ we have $a=\frac{196}{45}$. The values of $a$ 
at the other fixed points can be calculated numerically and
are reported in the last column of Table 2.

Since FP$_2$ and FP$_3$ have a unique irrelevant direction,
there is only one RG trajectory leaving these fixed points,
that can be integrated numerically
in the direction of increasing $t=\log k$ and ends up (in the UV)
at another fixed point.
In this way we have found an RG trajectory that goes from 
FP$_1$ to FP$_3$ and one that goes from FP$_4$ to FP$_2$.
The value of $a$ decreases along these trajectories,
in accordance with the theorem.
On the other hand, all the fixed points with $\lambda=0$
have very similar values of $a$
and there is a trajectory that goes from FP$_4$ to another
fixed point with $\lambda=0$ and a slightly larger value of $a$,
in contradiction to the theorem.
Since it is doubtful that these additional fixed points do exist,
the meaning of this result is not very clear,
and will have to be investigated more carefully in the future.

\subsection{Spectrum}

The appearance of several non-trivial fixed points is 
not a novelty in this kind of calculations.
Several of these are likely to be spurious,
but we do not see any reasons
why FP$_1$ or FP$_2$ should be rejected {\it a priori},
or to prefer one over the other.
Regarding the spectrum, we recall that in order to avoid tachyons
in the expansion around flat space,
the action for gravity in Lorentzian signature\footnote{we use the Lorentzian signature $-+++$.}
must have
a negative Weyl squared term and a positive $R^2$ term.
A naive Wick rotation of the linearized action around flat space
leads to a Lorentzian action that only differs from the
Euclidean one by an overall sign.
Therefore, FP$_2$ has the correct signs to avoid tachyons.
Although this is not sufficient to guarantee a healthy theory,
it gives us some more room in the search of one.
\\

Note added: After this paper was submitted to the journal the work referred  to in footnote 4 has appeared on the arXiv \cite{Kluth:2020bdv}.

%%%%%%%%%%%%%%%%%%%%%%%%%
\section*{Acknowledgment}

We would like to thank Dario Benedetti, Taichiro Kugo, Frank Saueressig and Omar Zanusso for valuable discussions.
This work was supported in part by the Grant-in-Aid for Scientific Research Fund of the JSPS (C) No. 16K05331.

%%%%%%%%%%%%%%%%%%%%%%%%%%%%%%%%%

\end{document}